# An Algorithm and Software for Establishing Heterogeneous Parking Prices

Nir Fulman, Itzhak Benenson

Geosimulation Lab, Department of Geography and Human Environment,

Tel Aviv University, Israel

nirfulma@post.tau.ac.il, bennya@post.tau.ac.il

Abstract

Parking prices in cities are uniform over large areas and do not reflect spatially heterogeneous parking supply and demand. Underpricing results in high parking occupancy in the subareas where the demand exceeds supply and long search for the vacant parking, whereas overpricing leads to low occupancy and hampered economic vitality. We present Nearest Pocket for Prices Algorithm (NPPA), a spatially explicit algorithm for establishing on-and off-street parking prices that guarantee a predetermined uniform level of occupation over the entire parking space. We apply NPPA for establishing heterogeneous parking prices that guarantee 90% parking occupancy in the Israeli city of Bat Yam.

# 1. From predefined and constant to adaptive and heterogeneous parking prices

Parking demand is defined by the destinations' capacity and attractiveness for drivers' activities and is essentially heterogeneous in space and in time. The price of parking, and especially of curb parking, does not reflect spatial-temporal variation of demand. Typically, it is uniform in space and is established for years over large urban areas. As a result, the parking supply-demand-prices triple is often unbalanced (Arnott & Inci, 2006; Shoup 2011). The problem is salient for both curb parking and off-street parking facilities. Underpriced parking in areas of high demand results in close to 100% occupation, long cruising time, and, indirectly, in traffic congestion and pollution. Overpricing results in under-occupancy and excessive parking in adjacent cheaper areas and is disadvantageous for the local economic activities.

## 1.1. From modeling parking choice to parking policy

Various economic models examine drivers' choice between parking lots and curb parking (Arnott, 2006; Anderson & de Palma (2004, 2007); Inci & Lindsey, 2014) and aim at optimizing curbside and garage parking pricing (See Inci, 2015 for a comprehensive review). The research that focuses on the negative effects of cruising for parking (Arnott & Inci, 2006; Shoup, 2006) claims that cruising is the result of underpriced parking prices and suggests increasing them to eliminate the problem. Another group of works links parking prices to congestion by embedding parking into traffic models (Arnott et al., 1991; Zhang et al., 2011; Fosgerau & de Palma, 2013; Verhoef et al., 1995). Although these models identify policy components that optimize average parking utilization, they are aggregative and pay limited attention to the spatial aspects of the parking process. As a result, it is difficult to translate their conclusions on parking pricing policy into practical recommendation.

The effects of potentially influencing factors on drivers' choice of parking types and locations are studied, mostly, with the discrete choice models, based on stated and revealed preferences data. Parking prices and distance from parking place to destination are considered as two major factors (Zhang & Zhu, 2016), and the increase in each of them negatively affects the probability to park. For example, Hensher & King (2001) estimate elasticity of parking demand with respect to parking price, walking distance and other properties. They find that a 1% increase in hourly parking rates results in a reduction in the probability to park of between 0.5% and 1%, and that a similar increase in walking time decreases the probability by 0.1% to 0.7%.

Possible influence of driver characteristics, including driver's income, on parking behavior are not always found significant or of essential importance (Simićević, 2013, Zhang & Zhu, 2016). However, this may be the result of respondents' reluctance to answer honestly regarding their income or other attributes (Shiftan & Burd-Eden, 2001). Recently Ibeas et al. (2014) examined parking behavior in a coastal town in Spain through an MMNL model and have demonstrated that the perception of parking charges is fairly heterogeneous, and depends both on the drivers' income levels and whether or not they are local residents.

## 1.2. The idea of adaptive parking prices

The starting point for establishing parking prices is Donald Shoup's (2006) idea that in order to eliminate cruising, on-street parking prices should be set to preserve parking occupancy at under 85% on every street link. This rule-of-thumb has recently reached practitioners and stakeholders, and several cities around the world have initiated pilot projects of local adjustment of curb-parking prices to demand. The cities of Los Angeles and San Francisco operate a demand-responsive parking policy that varies prices by street segments (LADOT, 2016; SFMTA, 2016), while in Calgary and Seattle prices are differentiated by areas of different sizes (CPA, 2011; SDOT, 2016). Typically, in these projects parking fees for the chosen spatial units are updated repetitively, once in a period of time, until occupancy rate falls within the range of around 60 – 80%.

The San Francisco project, called SF*park*, attracted special attention and was reviewed by a few scholars who concluded that it achieved its goals. Pierce & Shoup (2013) calculated the price elasticities of curb parking occupancy for each of the 6 price adjustments that were made between August 2011 and 2012. For each price change, they compared the old and new fee and average occupancy, as proxy for demand, in each street segment. They found that although elasticity varies by time of day and location, on average it is negative and occupancy levels generally move towards the target range for both under- and over-occupied blocks. Millard-Ball et al. (2014) Simulated cruising for parking and driver arrival rates using a Markovian queue model calibrated with data from SF*park* parking sensors. They demonstrate that on average, a rate change brings a street segment 0.1-0.2% closer to the 60-80% target and reduces cruising by more than 5%.

Yet demand-responsive parking is not free of shortcomings. First, in order to estimate parking occupancy, the projects employed expensive sensors. As a result, the pilot projects of San Francisco and Los Angeles cost millions of dollars to set up and operate. Moreover, the projects require adjusting the prices iteratively until the target occupancies are achieved, which may lead to frustration on the side of drivers.

We claim that parking prices that preserve a predetermined level of occupation can be established by using spatially explicit demand-supply models and present a novel algorithm for that. In what follows we limit ourselves to the basic factors that determine drivers' behavior: walking distance to destination, parking price and personal income. The proposed algorithm extends the algorithm proposed by (Levi & Benenson, 2015) for estimating spatially explicit demand/supply ratio. The equilibrium price pattern is estimated based on high-resolution GIS layers of roads, parking lots and buildings that are widely available at the municipal levels. The proposed algorithm is implemented with the freely available PARKFIT2 application and is freely available at https://www.researchgate.net/profile/Nir_Fulman.

## 2. Nearest Pocket for Parking algorithm
### 2.1. Initial settings

Nearest Pocket for Parking algorithm (NPPA) considers demand at a resolution of a separate building and parking facilities at resolution of a separate parking place and parking lot. These high-resolution data are easily available as standard components of municipal GIS: A layer of buildings with height attribute that can serve for estimating a number of apartments or offices that define parking demand and two layers of parking facilities, that of the street links (preferably with parking permissions) and of parking lots (with the attributes of lot area and number of floor or, preferably, lot's total capacity), can serve for estimating parking supply. In case the information on curb parking permissions is unavailable, parking places for parallel parking can be constructed automatically, 5 meters apart from each other on both sides of the two-way street link and on the right side of the one-way link with a necessary gap from a junction. Information on possible perpendicular parking and on parking permissions and restrictions can easily be accounted for. Similarly, if the capacity of the parking lot is unavailable, the number of places of the parking garage floor can be obtained as its total area divided into (5 + 2)*2 m rectangles, where additional 2 m of length reflects the in-lot space for maneuver (Figure 1).

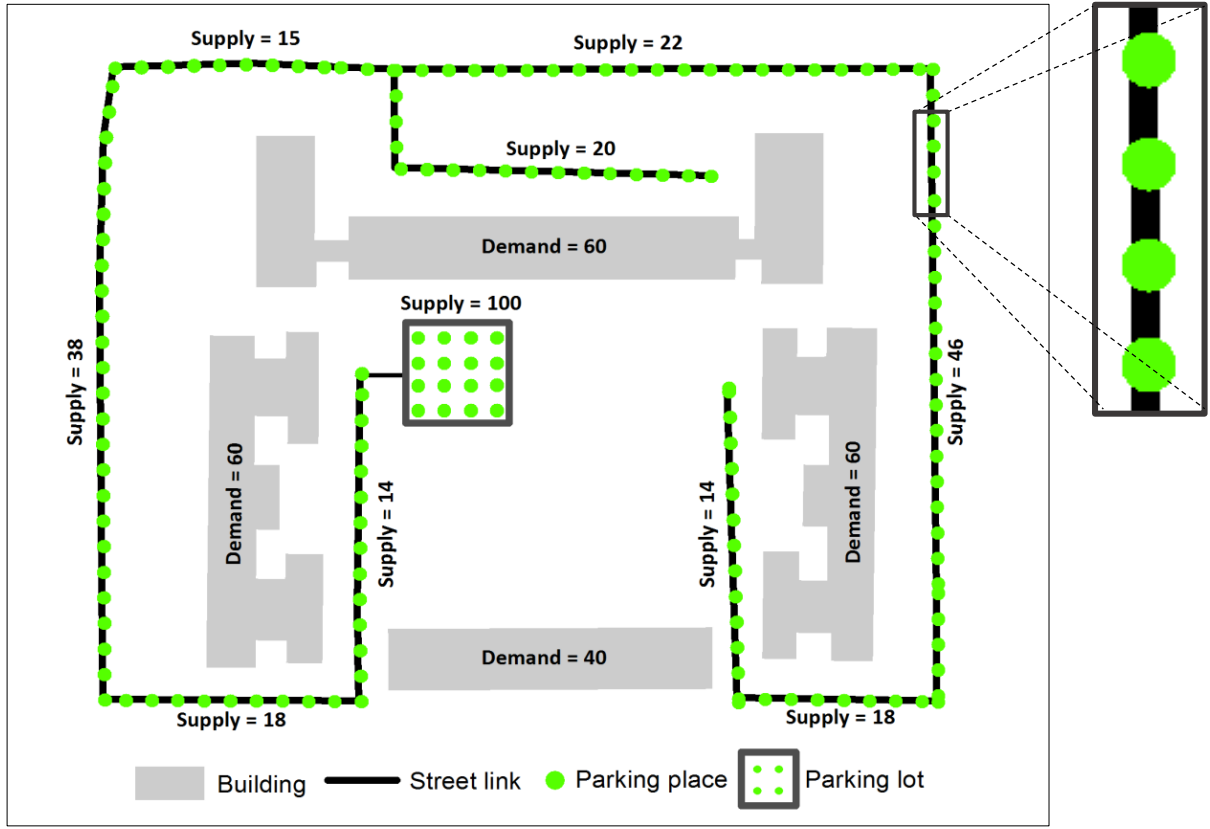

Figure 1. GIS layers for establishing heterogeneous parking price pattern: The layer of buildings that reflects the demand and the layers of parking lots and street links that reflect the supply. In case no other information is available, curb and lot parking places can be constructed automatically.

Street link and parking lot are typical minimal units for establishing parking prices. Larger units can also be considered. In case of curb parking, several connected street links, or all links within an urban neighborhood can be considered as a unit for establishing parking price. The goal of NPPA is to establish a pattern of parking prices that preserves the occupancy $O_{unit}$ of each parking unit below the threshold level $O_{th}$, $O_{unit} \leq O_{th}$.

### 2.2. The tradeoff between parking price and walking distance to destination

We assume that each driver c is characterized by its economic status and aims at parking as close as possible to its destination. The economic status defines driver's *minimal perceived price* (MPP) $w_{c,mpp}$ of the "best" that is, the nearest to the entrance to a destination, parking place. Below we exploit $w_{c,mpp}$ as a proxy of driver's economic status. Let the price of the best parking place be denoted as $F_p$. Let us denote $A_{c,p}(d)$ attractiveness of a parking place p and a distance d from the driver's c destination. Our major assumptions are as follows:

- Attractiveness $A_{c,p}(d)$ of a parking place p decreases with the distance d between the vacant and best possible parking place as $1/d^{\alpha}$, $A_{c,p}(d) \sim 1/d^{\alpha}$
- Attractiveness of a parking spot p for a driver c depends on p's price only when the $w_{c,mpp} < F_p$.

We combine these assumptions into the following formula for dependence of the $A_{c,p}(d)$ on $w_{c,mpp}$, $F_p$ and d:

$$A_{c,p}(d) = \min(1, w_{c,mpp}/F_p)/d^{\alpha} \qquad (1)$$

In what follows we consider sub-linear decrease of attractiveness with distance that is $\alpha < 1$.

We assume that there exists a maximal acceptable walking distance $d_{max}$ between a parking place and a destination and for $d > d_{max}$ $A_{c,p}(d) = 0$, and measure distance d in units of car length (5 m below). We additionally assume that for all drivers, $w_{c,mpp}$ is above some non-zero $w_{minmpp}$, and that for $F_p = 0$ $A_{c,p}(d) = 1/d^{\alpha}$. In numeric experiments below we set $\alpha = 0.5$ and $d_{max} = 100$ (500 m).

In what follows we consider two ways of partitioning urban parking space into units of uniform price: (1) each street link is a unit and (2) parking unit consists of all links within a traffic analysis zone (TAZ, Figure 7a). In all simulations below distance between parking spot and destination is the walking distance between them via the street network.

### 2.3. The Nearest Pocket Algorithm(NPA)

The NPPA extends the algorithm proposed by Levy and Benenson in their PARKFIT model (Levy & Benenson, 2015). As an initial step their algorithm is extended for estimating which parking units will be occupied at a level of $O_{th}$ in case prices are set to zero (Figure 2). Let for each destination k = 1, 2, 3, ..., K the number of drivers for whom k is the destination be $n_k$. The steps of the NPA algorithm are as follows:

(1) Build the list of all (driver, destination) pairs of length $n_1 + n_2 + n_3 + ... + n_K$) and randomly reorder it;

(2) Assign parking places to drivers according to the position in the list - m-th driver $c_m$ is assigned the nearest to $c_m$'s destination yet vacant parking place on a parking unit u where $O_u < O_{th}$; if there are no vacant spots in such units at a distance below $d_{max}$ from the $c_m$'s destination ignore the driver;

After assignment is completed, units u in which $O_u = O_{th}$ are the candidates to become overly occupied and their prices should be increased.

In what follows, we refer to this extension as Nearest Pocket Algorithm (NPA), and to the full algorithm presented as Nearest Pocket for Prices Algorithm (NPPA). To continue the model line, we call the application that implements NPPA as PARKFIT2.

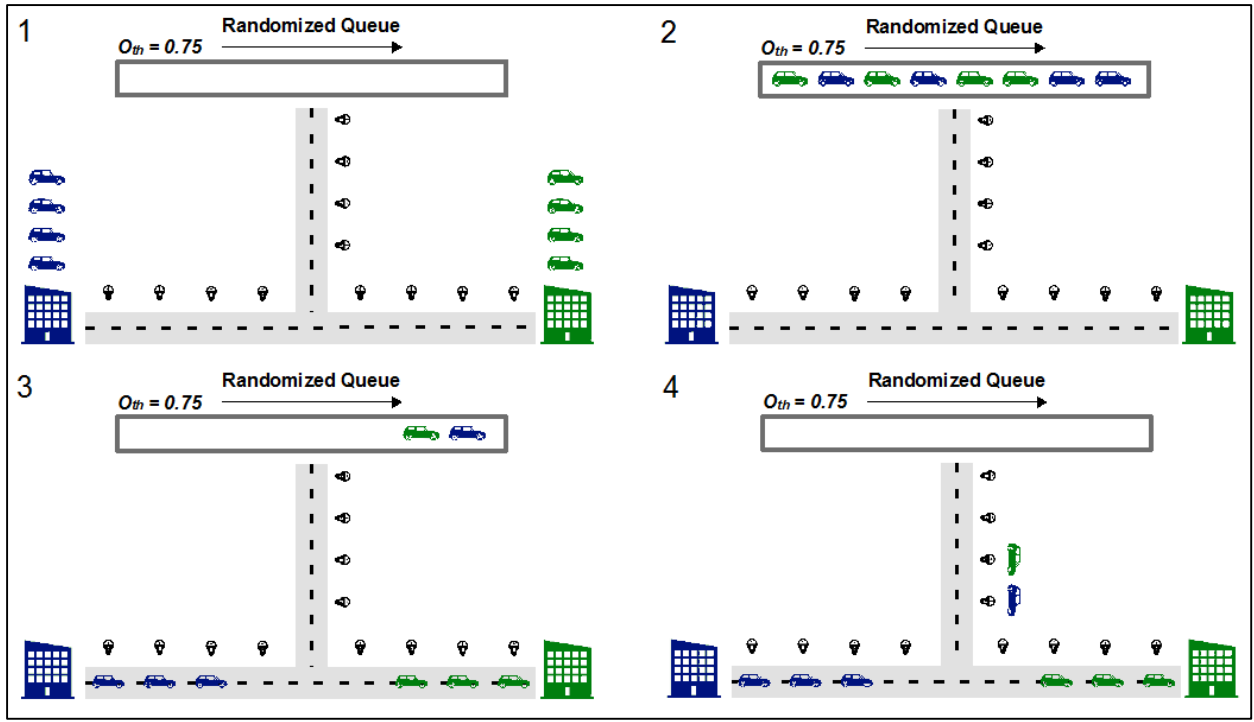

Figure 2: An illustration of NPA. (1) 4 Blue and 4 Green Drivers aim to park near destination of their color, occupancy threshold $O_{th}$ is set to 0.75; (2) List of (driver, destination) pairs is constructed and randomly reordered; (3) 6 First drivers in a queue are parked at the places that are nearest to their destinations. As result the bottom street segments are at 75% occupancy; (4) Remaining 2 drivers are forced to park on the third link and its occupancy reaches 50%. Only the bottom links are candidates for price increase.

### 2.4. Nearest Pocket for Prices Algorithm (NPPA)

Let $O_{th}$ be the threshold occupation rate. NPPA consists of two stages – establishment of initial price for every parking unit and convergence to an equilibrium distribution of prices, by parking units (Figure 3).

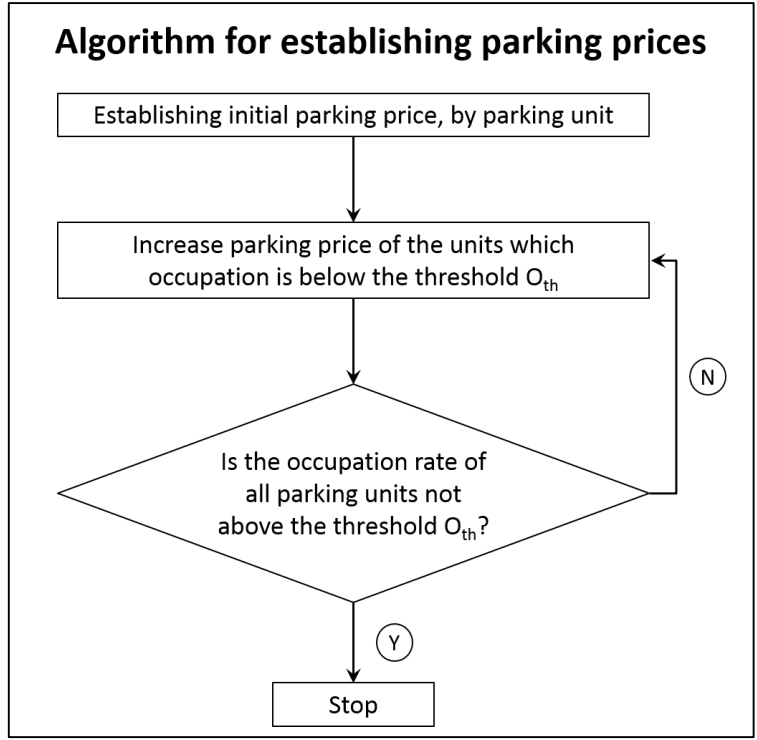

Figure 3: Major steps of the NPPA algorithm.

### 2.4.1. Stage 1, Establishing initial parking prices

(a) Apply NPA for $O_{th} = 1$.

(b) Let the number of parking spots at a unit u be $s_u$, the number of occupied places on u be $s_{u,occupied}$ and $s_{u,occupied} / s_u = O_u > O_{th}$. Build list $L_u$ of occupied parking spots on u and sort them in descending order of $w_{c,mpp}$ of the drivers that parked on it.

(c) Consider place $s_{th} = INT[s_u*O_{th}] + 1$ in $L_u$ and the widest symmetric neighborhood around it that is, all places between $s_{th} - (s_{u,occupied} - s_{th}) = 2*s_{th} - s_{u,occupied}$ and $s_{u,occupied}$; Calculate and store average $w_{c,mpp}$ of drivers parked in these places as an estimate of initial parking price $F_u$ on u

(d) Repeat stages (b) - (g) M times. In case the set of $F_u$ values is not empty, set initial parking price $F_{u,initial}$ as an average of the $F_u$ values; Otherwise set $F_{u,initial} = F_{u,min} = \min_c\{w_{c,mpp}\}$.

### 2.4.2. Stage 2, Iterative convergence to equilibrium price pattern

At the second stage of the algorithm, the price $F_u$ of excessively occupied parking unit u is increased and this causes the decrease of the occupation rates there. When the pric`e of parking at u reaches the level that guarantees that an average occupation rate there does not exceed $O_{th}$, the price of parking there stops growing.

At this stage of the algorithm we have to consider the case where the demand/supply ratio for the entire area is greater than the threshold occupation rate $O_{th}$. That is, no matter what are parking prices it is impossible to supply parking to all drivers who want to park and preserve the desired occupation rate. To resolve the problem we allow some of the drivers to skip parking in this round of parking search in the area. Formally, we assume that a driver c will skip parking with a probability $g_c(A_{c,best}) > 0$ if the attractiveness $A_{c,best}$ of the best available parking spot at a distance below $d_{max}$ from c's destination is under the threshold of attractiveness $A_{th}$. We formalize the dependence of $g_c(A_{c,best})$ on $A_{c,best}$ as follows:

$$g_c(A_{best}) = \begin{cases} 0 & \text{if } A_{best} > A_{th}, \\ 1 - \exp(\gamma*(1 - A_{th}/A_{best})) & \text{if } A_{best} \leq A_{th} \end{cases} \quad (2)$$

where $\gamma$ is a parameter.

In computational experiments below we employ, in addition to $\alpha = 0.5$ and $d_{max} = 100 = 500m$, the values of $A_{th} = 0.1$ and $\gamma = 0.1$. The value of $A_{th} = 0.1$ is chosen equal to the attractiveness of a parking place at a distance of 500m in case of parking prices below driver's MPP. For comparison, if $w_{mpp}/F_u = 0.5$ that is, the price of a parking place is relatively high for a driver, the value of $A_{th} = 0.1$ is achieved, for $\alpha = 0.5$, at a distance of 125 m.

The second stage of the algorithm differs from the first one in three respects:
- The attractiveness of each parking place for a given driver is estimated according to the formula (1) *accounting for the price of a parking place*.
- A driver may decide to skip parking entirely, according to (2).
- Prices of overly occupied parking units are iteratively increased until their average occupation rates decrease to $O_{th}$.

Formally algorithms steps at second stage are as follows:
(a) Apply NPA for $O_{th} = 100\%$, use parking spot attractiveness values calculated according to (1) with current prices $F_u$
(b) For each parking unit u, calculate and store the average occupancy $O_{M,ave}$ over M iterations of (a)
(c) If $O_{M,ave} > O_{th}$, increase parking price from $F_u$ to $F_u(1 + x)$

The value of x is a parameter of an algorithm and, evidently, the higher is x the faster will be the convergence of the pattern to the equilibrium one. However, the study of the algorithm shows that for high values of x, in iterations, the price of a link can fluctuate and not converge. Below we apply the value of x = 0.05 for which, empirically, fluctuations are not observed. For x = 0.05, in all experiments, the number of model iterations necessary for convergence to the equilibrium pattern was less than 60.

## 3. Study of the algorithm

We study the algorithm in two case studies. The first is an abstract grid city where we investigate the basic properties of the algorithm. In the second study we establish equilibrium parking prices for the Israeli city of Bat Yam. In both cases we consider night parking: the cars enter the area, park and do not leave. The price is set for the entire night. Unless noted otherwise, in all simulations below we consider the following set of parameters: $O_{th} = 0.85$, $\alpha = 0.5$, $A_{th} = 0.1$, $\gamma = 0.1$, $w_{minmpp} = 1$. The value of the $w_{mpp}$ varies between different group of drivers considered in experiments, while for each group we assume that the CV of the $w_{mpp} = 20\%$. In all simulations below, the number of repetition M = 20.

### 3.1. Abstract scenarios

We consider a grid city that consists of 60 two-way streets links of 100 m length and 400 buildings (Figure 4). The central building (black in Figure 4) can differ from the rest of the buildings in terms of the number of drivers for whom these buildings are the destinations and in the MPP of these drivers. On-street parking is permitted at a distance 10 m from the junction and the length of parking place is 5 m. That is, each link contains 40 parking spots and the total number N of parking spots in the area is N = 2400. Below, D denotes total demand for parking spaces and W drivers' average MPP for parking at the spot nearest to their destination.

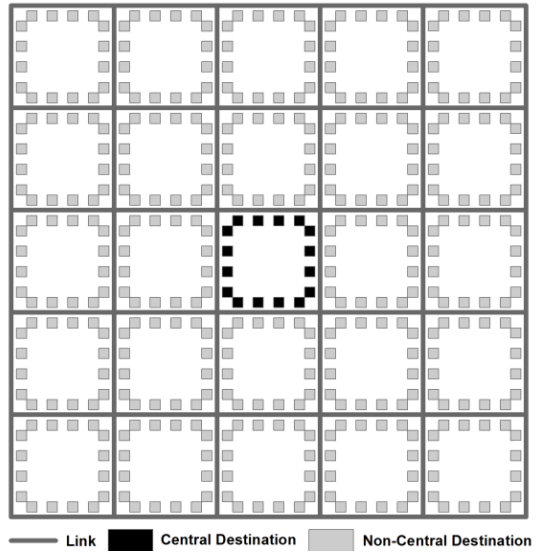

Figure 4: Links and destinations used in scenarios

Three scenarios that verify the steady parking price patterns are considered. The first two investigate the consequences of local excess of demand over supply and the third investigates the case where the total demand-to-supply ratio is higher than 1.

*Scenario A "high-rise buildings in the center of the city:"*

The goal is to estimate the steady price pattern in case of local demand/supply higher than one and homogeneous drivers' population. Average $w_{mpp}$ = 3 for all drivers. Demand for the 16 central destinations is set equal to 16 and total demand of all four buildings on one side of a street link is 56, 275% of parking supply on the same side. Parking demand of the rest of buildings is 4 that is, total demand of all buildings on one side of a street link is 4*4 = 16, 75% of the link's side parking supply.

*Scenario B "high-rise buildings for the rich in the center of the city"*

The goal is to estimate steady price pattern in case where both local demand/supply is higher than 1 and the drivers whose destinations are in the center are richer than the rest of the drivers. Spatial distribution of demand is set the same as in scenario A, but for drivers whose destination is in one of the central houses average $w_{mpp}$ = 5, while for the rest $w_{mpp}$ = 3.

Scenario C *"high-rise buildings for rich in the center of the city; total demand/supply ratio is above 1"*

The goal is to estimate the steady price pattern in case where the total demand/supply ratio is higher than 1, for the same conditions as in scenario B - local demand/supply in the center is higher than at the periphery and the drivers whose destinations are in the center are richer than the rest of the drivers. As in scenario B, for drivers whose destination is in one of the central houses average $w_{mpp}$ = 5, while for the rest $w_{mpp}$ = 3. The demand is increased 50% relative to scenario B: for 16 central destinations it is set to D = 24, while for the rest of the buildings it is D = 6. That is, the total demand D = 2688 and for $O_{th}$ = 0.85 at least 2688 – 2400 * 0.85 = 648 drivers that have to give up on parking in the area.

In what follows, we present scenarios results by the distance classes from the center of the city (Figure 5b).

In both scenarios A and B, the prices should be raised over all besides the boundary links (Figure 5c) and no drivers give up on parking. Model price pattern converges to a concentric hill (Figure 5d) and the competition for parking places in the overpopulated grid center leads to higher prices there, evidently higher in scenario B than in A and reflecting higher MPP of center's residents in scenario B. "Rich drivers" of scenario B push the rest of drivers away from the center of the city: Average walking distance from parking place to destination for the drivers whose destinations are in 16 centermost buildings decreases from 120 meters in scenario A to only 70 meters in B, while for the rest of drivers, average walk distance increases from 15 m in scenario A to 25 meters in scenario B.

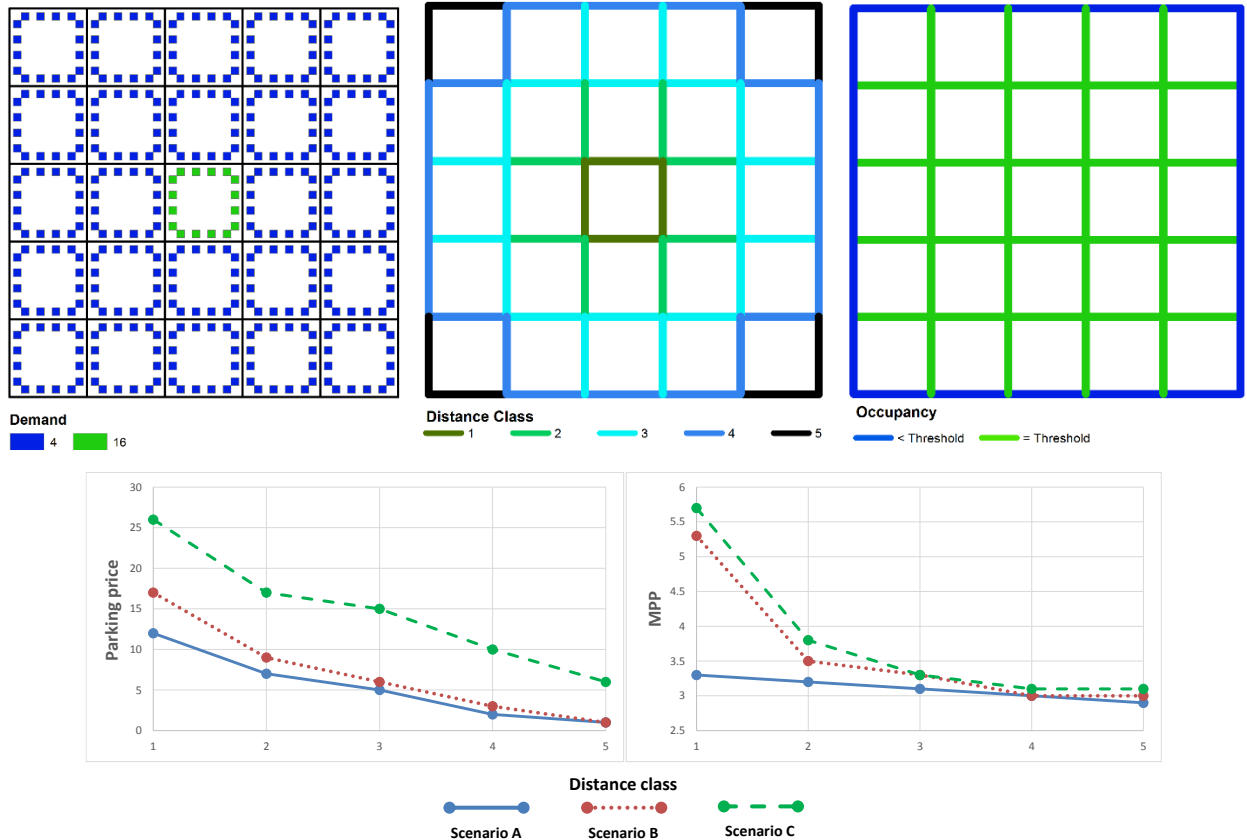

Figure 5: Distribution of demand for scenarios A, B and C (a); distance classes from the central block (b); links where the price should be increased according to NPA (Levy & Benenson, 2015) in scenario B (c); Parking prices, averaged by distance classes (d) Average MPP of drivers parked on street links, averaged by distance classes (e).

The number of drivers giving up on parking in scenario C varies between 745 and 772 that is, remains always higher than the theoretical minimum of 648. This happens because some of the costly and distant parking spots which attractiveness is far below $A_{th} = 0.1$ remain the only option for the driver. Steady prices in scenario C are higher than in B everywhere over the city (Figure 5d), while the distribution of the MPP of parked drivers is very close to that in scenario B (Figure 5e).

### 3.2. Sensitivity of equilibrium urban pattern to distance decay of the attractiveness

The decay of attractiveness with the distance is represented by the value of α in (1). To investigate sensitivity of the price pattern to the value of α we have varied it between 0.2 and 0.6 for scenario A above. To remind, $O_{th}$ = 0.85, the overall D/N ≈ 0.75, in 16 centermost D = 16, for the rest of destination D = 4, and average $w_{mpp}$ = 3 for all drivers with the CV = 20%. We characterize sensitivity to α by comparing prices decay with the distance from the central block. Higher α means larger difference between the attractiveness of the closer and more distant spots and with the growth of α, price pattern is less and less uniform (Figure 6).

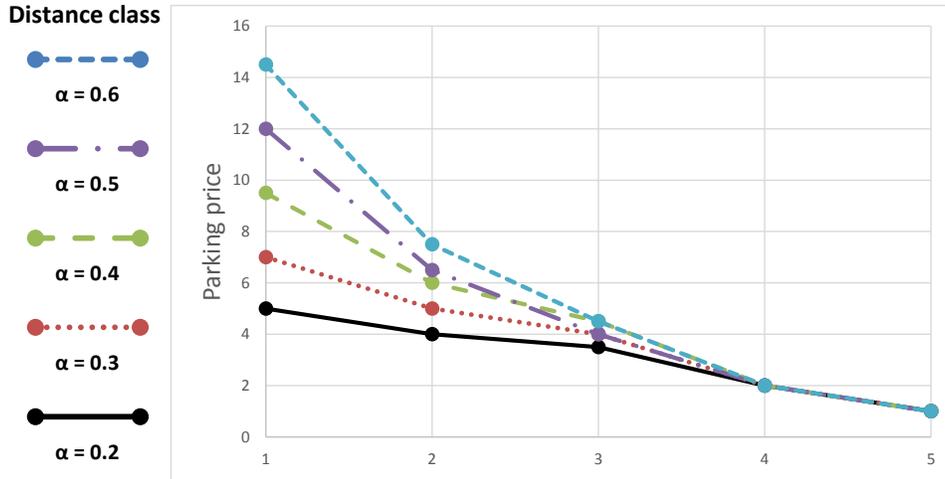

Figure 6: Averages of prices, by distance

**Establishing heterogeneous parking prices in the city of Bat Yam**

To investigate the algorithm in a real-world heterogeneous environment, we applied it for establishing overnight parking prices in the city of Bat Yam, Israel. In 2010, the population of Bat Yam was 130,000, total car ownership 35,000, total number of residential buildings 3,300 and apartments 51,000. The residential buildings in Bat Yam provide their tenants a total of 17,500 parking places, thus the ratio of drivers who seek parking to apartments is (35,000 − 17,500) / 51,000 ≈ 0.34. The Bat Yam GIS for the year 2010, supplied to us by the city's municipality, contains layers of streets, parking lots and buildings. We associate destinations with residential buildings and estimate the demand for parking in each as 0.34 times the number of apartments in the building.

Based on the GIS street layer, 27,000 curb parking places were constructed. In addition 1,500 places are available free of charge for the city's residents in its parking lots. The average overnight demand-to-supply ratio is thus (35,000 − 17,500) / (27,000 + 1,500) ≈ 0.61 cars/parking place. However, the distribution of demand and supply is highly heterogeneous. The estimates of both demand and supply concur with a field survey conducted in 2010 (Levy & Benenson, 2015).

Occupancy threshold is set to 92%, above which drivers' cruising time starts to grow (Levy et al., 2013). Based on the information of residents' economic status supplied per traffic analysis zone (TAZ, Figure 7a), MPP = 3 is assigned to the residents of the poorest statistical block. For the residents of the other blocks average MPP is set proportionally to the ratio of their economic status to the economic status of the residents of the poorest block, with a 20% coefficient of variation within the block.

The purpose of the Bat Yam case study is to explore a realistic scenario in terms of scale, heterogeneity of supply, demand and the level of drivers' income, as well as complexity of the street network. We thus ignore other factors that can influence parking choice of the Bat Yam's residents. This is, formulas (1), (2) and (3), are set up as in section 3.1. We have considered two partitions of the Bat Yam parking space into units. One is based on street links and parking lots and one on TAZ areas. That is in the former case the price of parking is defined for each street link and parking lot, while in the latter for all street links and parking lots in a TAZ.

### 3.3. Parking price pattern at resolution of street links and parking lots

Although the global demand/supply in Bat Yam is far below 1, the distribution of demand is highly heterogeneous and potential overnight parking occupancy estimated according to NPA is above 0.92 in over 2/3 of the city area (Figure 7b). That is why the prices of the equilibrium pattern obtained with the PARKFIT2 algorithm reach the level of 10 – 20 NIS in the central part of the city (Figure 7c). As can be seen, high heterogeneity of demand results in high heterogeneity of parking prices.

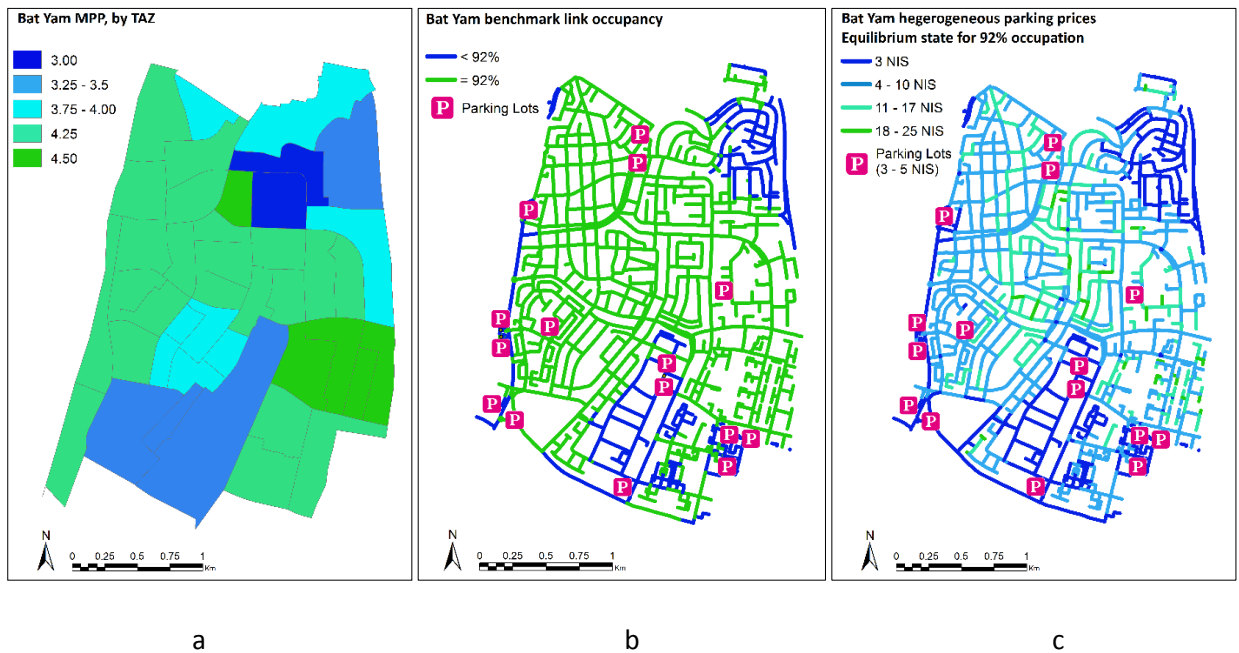

a　　　　　　　　　　　　　　b　　　　　　　　　　　　　　c

Figure 7: Bat Yam Income as translated into the MPP, by TAZ (a); parking demand by street segments (b); equilibrium parking prices, in NIS, for 92% occupation threshold estimated for the street links as parking units (c).

### 3.4. Establishing parking prices by TAZ

The number of parking places in a TAZ is, on average, about 700, roughly 50 times more than on a street link. That is why the price that is uniform over **all** parking places in TAZ may mask essential differences between the link-based demand/supply ratios there; nonetheless, the demand/supply yet essentially varies by TAZ. Figure 8a presents these ratios, while the other parts of Figure 8 present model outcomes: Figure 8b presents TAZ where the prices should be increased; Figure 8c NPPA outcome that is based on TAZ as parking units, and Figure 8d presents, for comparison, the average, by TAZ, prices that were established for Bat Yam based on the link as parking units (Figure 7c).

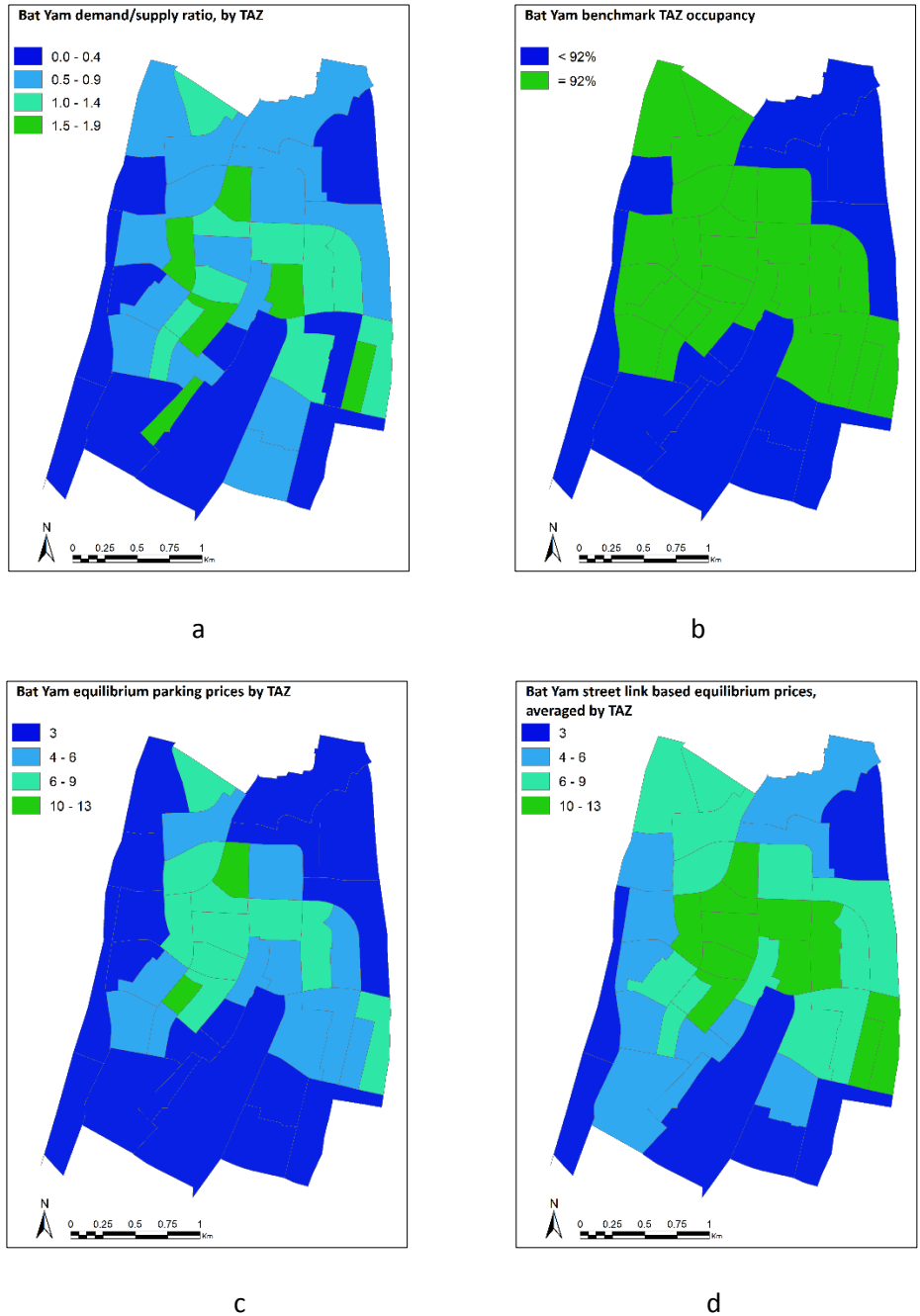

a
b
c
d

Figure 8: Demand/supply ratio, by TAZ (a); Parking demand by TAZ (b); parking price pattern for direct calculations at TAZ resolution (c); Link-based parking prices from the Figure 7c, averaged by TAZ (d).

While the maps in Figure 8c and 8d are qualitatively similar, peripheral prices in the TAZ-based model are lower. The phenomenon is a direct consequence of the low D/N value in the peripheral Bat Yam TAZ. In the TAZ-based version, parking prices for the peripheral TAZ are established accounting for numerous underused links, which overweight links with higher D/N ratio within this TAZ.

## 4. Discussion

Spatially explicit, high-resolution NPPA algorithm for establishing heterogeneous pattern of urban parking prices is proposed and implemented in PARKFIT2 software. The algorithm accounts for spatial distribution of parking demand and supply and for the distribution of minimal perceived price in the population. It exploits the advantages of demand-based pricing, but does not require street equipment for price adjustments. PARKFIT2 is applied for establishing heterogeneous parking prices in the Israeli city of Bat Yam.

### 4.1. Occupancy thresholds and the size of a parking unit

According to Shoup (2006), cities should aim at 85% parking occupation level, at which cruising behavior is minimized. Levy et al. (2013) show that this level can be raised to 92-93%. However in San Francisco's SFpark experiment, the upper bound of the target average occupancy range was set to 80%. The gap is not arbitrary. As Pierce and Shoup (2013) explain, if the bar is set too high, it is impossible to avoid reaching full parking capacity often, due to the stochastic variation in parking demand. The issue of stochastic variation in parking demand can be treated in PARKFIT2 by simulating the maximum expected demand, as is done in the Bat-Yam case study based on the capacity of destinations.

But variation in occupancy also stems from the stochastic arrival order of drivers who wish to park. Occupancy of smaller parking units is more influenced by this variation than larger ones, and they are more likely to become fully occupied. Due to this, maximum occupancy of units does not necessarily imply increased cruising time, and should not be considered when determining the occupancy threshold in PAKRFIT2.

More importantly, where variation of a single occupancy measurement is higher, the average is more influenced by the number of iterations M. Thus small units and insufficient number of iterations may lead to price fluctuations that result in higher prices and to equilibrium price patterns that differ greatly from each other. For this reason, parking units shouldn't be too small. In any case PARKFIT2 should be run with an increasing number of iterations, until the parking prices converge to similar equilibrium patterns. The number of iterations may be limited due to considerations of computational performance, in which case, it is possible to increase the unit size. However note that maximum prices are higher for smaller units, as they account for the spatial distribution of demand with higher resolution. This issue is pointed out in the comparison of price patterns established for street links and TAZ in Bat Yam (section 4).

### 4.2. Varying parking fees with respect to time

Time-varying parking prices are necessary for coping with variation in demand. Van Ommeren & Russo (2013) show that fixed prices are inferior to prices that vary by day of the week in reducing parking costs of a hospital in the Netherlands. In San Francisco's *SFpark* (2016) experiment, prices vary between weekdays and weekends, and also by time of day. Pierce & Shoup (2013) further suggest seasonal adjustments of prices to cope with periodic and predictable patterns of demand.

PARKFIT2 can be applied for establishing time-varying parking prices by forecasting and adjusting the demand D for parking in destinations. D should represent the maximum demand during a period of time for which prices are set. For long periods, the capacity of a destination can be taken as its maximum demand, as we have done for establishing residential night-time parking prices for Bat-Yam (section 4). Yet for shorter time bands, a more elaborate approach of demand estimation should be adopted. To account for cyclical variations (e.g. days of the week and holidays) and special occasions, demand can be simply increased or decreased according to expectation.

### 4.3. Future research

Our model study raises several issues that are relevant for urban parking policy: What parking units should be chosen for establishing parking prices – does the block level approach used in Los Angeles and San Francisco achieve lower cruising time than the zonal pricing implemented in Calgary and Seattle? How should parking prices vary in time? How should we preserve parking space for the weaker population groups? How can the current parking prices be substituted by heterogeneous ones? What is the seamless way of transition from the current to heterogeneous prices?

# 6. Appendix: Full Nearest Pocket for Prices algorithm

*Notation*

$A_{c,p}(d)$ - attractiveness of a parking place p for a driver c, whose minimal perceived price is $w_{c,mpp}$:

$$A_{c,p}(d) = \min(1, w_{c,mpp}/F_p)/d^\alpha \qquad (1)$$

where d is the distance between the parking spot p and c's destination d and $F_p$ is the parking price on p; we use $\alpha < 1$ to reflect a slow decrease in A with an increase in a walk distance;

$g_{c,p}$ - probability for a driver c to give up on parking at a spot p of attractiveness A:

$$g_c(A_{best}) = \begin{cases} 0 & \text{if } A_{best} > A_{th}, \\ 1 - \exp(\gamma*(1 - A_{th}/A_{best})) & \text{if } A_{best} \leq A_{th} \end{cases} \qquad (2)$$

where $A_{th}$, a threshold of attractiveness and γ are parameters;

$n_k$, k = 1, 2, …, K - demand for parking at a destination k, $n_{total} = n_1 + n_2 + n_3 + … + n_K$ – total demand;

$d_{max}$ - maximum acceptable walking distance between parking place and destination; $\min_c\{w_{c,mpp}\} > w_{minmpp} > 0$: $w_{minmpp}$ is the minimum of drivers' mpp, very low price that is below the price perceived by any driver.

*Stage 1: Establishing initial parking prices*

(a) Assign ID to each of $n_{total} = n_1 + n_2 + n_3 + … + n_K$ drivers
(b) Construct the list of triples (driver's ID, driver's $w_{mpp}$, driver's destination) of $n_{total}$ length and randomly reorder it
(c) Consider m-th triple in the list and assign m-th driver to the vacant parking spot that is nearest to m-th driver's destination, if such a spot exists within the maximum acceptable walking distance $d_{max}$. If all spots at a distance below or equal to $d_{max}$ are occupied, skip the triple
(d) Store the resulting quadruple (parking place ID, driver's ID, driver's $w_{mpp}$, driver's destination)
(e) Repeat (c) and (d) $n_{total}$ times
(f) Let for the unit u of parking capacity $s_u$ the number of occupied places is $s_{u,occupied}$ and $s_{u,occupied}/s_u = O_u > O_{th}$. Build list $L_u$ of all occupied parking spots sorted in descending order by the $w_{c,mpp}$ of drivers parked there
(g) Consider the place $s_{th} = INT[s_u*O_{th}] + 1$ in $L_u$ and the widest symmetric neighborhood around it that is, all places between $_{th} - (s_{u,occupied} - n_{th})$ and $s_{u,occupied}$. Calculate and store average MPP of drivers parked in these places as an estimate of initial parking price $F_u$ on u
(h) Repeat stages (b) - (g) M time. In case the set of the $F_u$ values is not empty, set initial parking price $F_{u,initial}$ as an average of the $F_u$-values. Otherwise set $F_{u,initial} = \min_c\{w_{c,mpp}\}$

*Stage 2: Iterative convergence to equilibrium price pattern*

(a) Randomly reorder the list of triples (driver's ID, driver's $w_{mpp}$, driver's destination)
(b) Consider m-th triple in the list and build the list of all vacant parking spot for the m-th driver within the maximum acceptable walking distance $d_{max}$. If this list is empty, skip the triple; otherwise order this list by the attractiveness of the parking spots in it for a driver, loop by the ordered list and, if $A_{c,p}(d) < A_{th}$ for a parking spot p, then skip this place with probability $g_{c,p}$. If a place is not skipped then park there and quit the loop
(c) Store the triple (parking place ID, driver's ID, driver's destination)
(d) Repeat (b) and (c) $n_{total}$ times
(e) Repeat (a) – (d) M times
(f) For each parking unit u, calculate the average occupancy $O_{M,ave}$ over M iterations
(g) If $O_{M,ave} > O_{th}$, increase parking price from $F_u$ to $F_u(1 + x)$, x is a parameter
(h) Repeat (a) to (f) until all prices remain unchanged in (f)